# Effect of pulse duration on the X – ray emission from Ar clusters in intense laser fields

Prigent Christophe, Lamour Emily, Rozet Jean - Pierre, and Vernhet Dominique
Université Pierre et Marie Curie - UPMC, UMR 7588, INSP, Campus Boucicaut, 140 rue de Lourmel, Paris, F-75015 France, EU and
CNRS, UMR 7588, INSP, Campus Boucicaut, 140 rue de Lourmel, Paris, F-75015 France, EU
Deiss Cornelia and Burgdörfer Joachim
Institute for Theoretical Physics, Vienna University of Technology, Vienna, A-1040 Austria, EU

We study experimentally and theoretically the production of characteristic K$\alpha$ X-rays during the interaction of intense infrared laser pulses with large (N~$10^4$ – $10^6$ atoms) argon clusters. We focus on the influence of laser intensity and pulse length on both the total X-ray yields and the charge-state distributions of the emitting cluster argon ions. An experimental optimization of the X-ray yield based on the setup geometry is presented and the role of the effective focal volume is investigated. Our theoretical model is based on a mean-field Monte-Carlo simulation and allows identifying the effective heating of a subensemble of electrons in strong fields as the origin of the observed X-ray emission. Well-controlled experimental conditions allow a quantitative bench marking of absolute X–ray yields as well as charge state distributions of ions having a K shell vacancy. The presence of an optimum pulse duration that maximizes the X-ray yield at constant laser energy is found to be the result of the competition between the single cluster dynamics and the number of clusters participating in the emission.

36.40 -c, 52.50.Jm, 52.38.Kd, 52.70.La

## I. INTRODUCTION

The unique properties of the interaction between strong laser pulses and clusters of atoms and / or molecules, has paved the way for a better understanding of the behavior of matter under intense electric perturbation. Particular features, such as production of energetic and highly charged ions (e.g. $Xe^{q+}$ charge states with q>10 and kinetic energies higher than 100 keV[1]), ejection of hot electrons (with energies ~ keV[2]) and emission of extreme UV and X–ray photons characteristic of inner-shell ionization of atoms [3,4,5], give evidence of a very efficient coupling of the laser field to matter. Large rare-gas clusters are finite systems which unite the advantages of gaseous and solid targets. On the one hand, the low *mean* atomic density ($n_{mean}$ ~ $10^{14}$ to few $10^{17}$ at./cm$^3$) of the cluster jet leads to a large distance between the clusters (of the order of micrometers), and thus allows for the study of the response of individual clusters in a well-defined laser field, undisturbed by propagation effects. On the other hand, since the *local* atomic density in the cluster is close to solid ($n_{local}$ ~ $10^{22}$ at./cm$^3$), the absorption of the laser pulse energy is locally as strong as in laser – solid interactions[6], and collective motion of quasi-free electrons amplified by high local electric fields may occur.

Stages of the dynamics of laser-cluster interaction can be summarized as follows: the atoms of the cluster are ionized by the incident laser pulse and a cold "nanoplasma" of solid density is formed. The quasi-free electrons are heated in the combined field of the laser and the surrounding particles. Electron-impact ionization produces higher charge states of cluster ions and inner-shell vacancies which are at the origin of characteristic X-ray radiation. As a fraction of the quasi-free electrons leave the cluster, a net positive charge is left behind and the cluster ions begin to expand before the cluster disintegrates completely in a Coulomb explosion. Due to the large inertia of the ions, the cluster expansion starts to play a role on the time-scale of several 10 to 100fs. This allows the interplay between the electronic and ionic dynamics to be effectively studied and controlled by varying the pulse duration of the driving laser pulse.

Several experimental studies have shown that energy absorption[7], production of highly charged and energetic ions[8,9], and photon emission[10,11,12] may exhibit non-monotonous dependencies on the pulse duration. So far, these experimental results have been compared [9,11,13] mostly to the nanoplasma model[14] where energy absorption occurs at the "plasma (or cluster) resonance", i.e. when the density-dependent plasma frequency in the cluster matches the laser frequency. Variation of the pulse duration should then allow for an optimal timing of the resonance[7,10,13]. For very short laser pulses, the pulse passes before the cluster has enough time to expand to the critical density where the resonance condition is satisfied, and consequently only limited energy absorption is expected. For a longer pulse duration the resonance condition is reached when the electric laser field is close to its maximum, leading to a much more efficient absorption. Finally, if the pulse gets too long, the major part of the laser pulse interacts with a greatly expanded cluster and leads again to low levels of energy absorption. Self-consistent 2D particle-in-cell (PIC) simulations of large clusters[15] have also stressed the influence of the pulse length on the heating of the electronic ensemble and traced a drop of the energy absorption at long pulse durations to the early disintegration of the cluster.

We present experimental results for the influence of the laser pulse duration on the keV X-ray yields from large ($N > 10^4$ atoms) argon clusters. Assisted by our mean-field particle simulations we aim to shed light on the temporal competition between the different processes occurring during the laser-cluster interaction which are still controversially debated[16], such as (multi-)ionization of the cluster ions, electron energy gain and cluster expansion. We find that not only the number of X-ray photons emitted per cluster but also the number of emitting clusters depends on the laser pulse duration. In light of these findings we propose an explanation for the observed maxima in the X-ray yield unrelated to nanoplasma resonances.

The paper is organized as follows: in section 2 we describe the experimental and theoretical methods. Section 3 focuses on the intensity dependence of the X-ray yield and the ionic charge state distributions for a given laser pulse duration. An experimental optimization of the X-ray yield based on the setup geometry is presented. The theoretical treatment gives insight into the single cluster dynamics and analyses in particular the effective heating of electrons up to keV energies during the interaction. The influence of the pulse duration on the X-ray emission is subject of section 4. The dependence of the yield on the laser intensity for different pulse durations gives insight into the evolution of the intensity threshold for X-ray production. Comparisons between experimental and theoretical charge state distributions for different pulse durations are performed. The dependence of the absolute X-ray yield on the pulse duration is then studied at fixed pulse energy and the origin of an optimum pulse duration is investigated. Finally, we present results for the X-ray yield as a function of the pulse duration at constant laser intensity.

## II. EXPERIMENTAL AND THEORETICAL METHODS

### A) Experimental Method

Whereas the spectroscopy of the emitted ions and electrons extracts information from the system a few microseconds after the femtosecond laser pulse and the cluster disintegration, X-ray spectroscopy allows performing measurements on an ultra short time scale down to a few tens of femtoseconds[17]. Ions with inner-shell vacancies contribute to X-ray emission, which gives access to the dynamical properties of the irradiated cluster on a time-scale smaller or comparable to that of the laser pulse duration. Furthermore, as the inner-shell vacancies are predominantly produced by electron-impact ionization[4,18], X-ray spectroscopy can provide deeper insight into the electron dynamics which is key to a detailed understanding of laser-cluster interaction.

Details on the experimental set up and on the methods used to control the different parameters acting in the laser-cluster dynamics have already been described in previous papers[19] (and references therein). We focus on those experimental characteristics which are of importance in the discussion of the results presented in this paper. Experiments have been performed on the LUCA facility (French acronym for Ultra Short Tunable Laser, CEA) at Saclay, where the laser field is generated with a Ti:sapphire laser system delivering pulses of duration down to $\tau = 50$ fs at full width half maximum (FWHM), centered at $\lambda=800$ nm with a repetition rate of 20 Hz. The laser beam diameter is approximately 50 mm before focusing and the maximum pulse energy available in the interaction zone is 50 mJ. The laser light is focused by a 480 mm focal lens leading to a beam waist, i.e. a full width at $1/e^2$ in intensity ($2.w_0$), of $31\pm1$ μm (corresponding to a Rayleigh distance $z_R$ of 0.95 mm) and maximum peak intensities $I_{peak}$ ~ $10^{17}$ W/cm². The $I_{peak}$ values are determined by imaging the focal point on a CCD camera and from systematic measurements of the effective energy contributing to the peak intensity and of the laser pulse duration via second-order autocorrelation techniques that take into account the spatio-temporal aberrations induced by the lens (a maximum broadening of 18 % is found at 50 fs while it drops down to less than 1% for a 200 fs pulse). Control of the laser parameters leads to an accuracy better than ± 20% on $I_{peak}$ values ranging from $10^{14}$ to $10^{17}$ W/cm².

The clusters are produced by condensation of gas flowing at high pressure through a conical nozzle[20]. The mean atomic density is proportional to the backing pressure $P_0$, and a mean cluster-size scaling roughly with $P_0^{1.8}$ is found[21]. The mean size <N> of the rare gas clusters is obtained combining the well known Hagena semi – empirical laws[22] and a compilation of recent experimental results[23] in the size range studied here *i.e.* from $10^4$ to $10^6$ atoms per cluster corresponding to a radius between 5 nm and 21 nm for argon clusters. In the present experimental conditions, the mean atomic density is of the order of $10^{17}$ at/cm³ and consequently the distance between clusters is ~ 0.7 μm. The nozzle is mounted on a solenoid pulsed valve, with an opening duration of

500 µs, operated at a repetition rate from 20 Hz to 1 Hz and synchronized to the laser source. The spatial and temporal overlap between the laser light and the cluster jet has been carefully checked and optimized[19] ensuring reproducible experimental conditions "shot-by-shot", *i.e.* a well-defined cluster bunch interacting with every single laser shot. By spatial X-ray optimization measurements, we have found the cluster beam FWHM to be of the order of several mm[19] with a rather flat cluster density profile in agreement with simulations of Boldarev *et. al.*[24] and measurements in ref 25.

The emitted X-rays are analyzed by two semiconductor detectors and a crystal spectrometer. Taking full advantages of the "shot-by-shot" regime, X- rays are recorded in coincidence with the laser pulse which reduces dramatically the background contribution (by around 4 orders of magnitude). A description of these spectrometers and of the determination of the absolute X-ray yields has previously been given[3,4,17,21]. Briefly, one of the semiconductor detectors is used in the pile-up mode, while the other one records single-photon spectra in order to evaluate the energy of emitted photons. This method makes it possible to track the variation of absolute X-ray yields over more than 5 orders of magnitude with uncertainties ranging from 15 to 30%. State-resolved measurements are performed using a high-resolution high-transmission Bragg crystal spectrometer, equipped with a flat mosaic graphite crystal (HOPG) and a large home-made position sensitive detector working in the photon counting mode. A typical efficiency of $2\times10^{-6}$ and a resolving power of 2000 allow determining the charge state distribution of the ions emitting X-rays (i.e. ions with inner-shell vacancies). In the case of argon clusters, the changes in intensity of the characteristic K X-rays (namely 2p→1s and 3p→1s deexcitations) from $Ar^{12+}$ to $Ar^{16+}$ ions[19] can be followed as a function of individual parameters (pulse duration, laser intensity, cluster size…). It is worth mentioning that the simultaneous measurements of absolute X-ray yields by the different detection devices (semiconductor detectors and crystal spectrometer) are in complete agreement with each other.

### B) Simulation method

In order to shed light on the experimental findings, we have developed a computer simulation of the laser-cluster interaction[18,26]. The large size of the clusters, the long pulse durations and the multitude of mechanisms at play provide a major challenge for a theoretical description and make a simplified approach inevitable. We opt for a mean-field approach in which many-particle effects are included via Monte-Carlo events. For the largest clusters ($N > 10^5$ atoms), we furthermore employ a test particle discretization, i.e. we solve the equations of motion only for a representative fraction of the ensemble of particles, which is determined by computational capabilities.

The mean field is evaluated at each time step $\Delta t = 0.01$ fs by solving the Poisson equation[27] on a cylindrical mesh. The mesh divides the simulation volume into $256 \times 512$ cells in (R,z), making use of the rotational symmetry around the polarization axis z of the laser. The cells are equally spaced in z-direction, while the radial steps in R are chosen in order to achieve equal volume of all cells. Outside the simulation volume, the mean field is assumed to be determined by the net charge and dipole moment inside the simulation volume. The particles move in the combined field of the mean field and the laser field $\mathbf{F_L}$. The laser wavelength being larger than the cluster radius, the laser field can be considered uniform on the scale of a single cluster:

$$\vec{F}_L(t) = F_0 \sin(\omega t) \cdot \sin^2\left(\frac{\pi \cdot t}{2\sqrt{2} \cdot \tau}\right) \vec{z} \tag{1}$$

with $F_0$ the maximum electric field, $\omega$ the laser frequency and $\tau$ the pulse duration. Particle-particle interactions such as elastic electron-ion scattering, electron-impact ionization, electron-impact excitation, are included by stochastic changes of the particle momentum controlled by the corresponding cross-sections and the local particle densities[18,28]. The cross-sections for elastic electron-ion scattering are calculated by partial wave-analysis of parametrized Hartree-Fock potentials[18,29,30], the impact ionization and excitation cross-sections are estimated with the Lotz formula[31] from the charge-state dependent ionic binding energies[32]. Electron-electron collisions are included following the proposal of Nanbu[33] for the cumulative treatment of small-angle collisions in plasmas: in each time-step, electrons from the same mesh cell are paired up randomly and undergo a binary collision that represents all the successive small-angle collisions that occurred during $\Delta t$.

The present simulation technique cannot resolve the mean interaction potential between electrons and ions on a length scale smaller than the cell size which is of the order of 0.5 nm (or 10 a.u.). It should be noted that this limitation does not affect collisional interactions which are accounted for by the stochastic event-by-event sampling. It influences, however the mean ion-electron and thus mean ion-ion interaction responsible for cluster expansion. The shielding of forces acting on an ion surrounded by (slow) quasi-free electrons is therefore not fully included in our simulation. Two limiting cases are considered in order to estimate its overall effect: i) the fully shielded case where all surrounding electrons contribute to the shielding and the effective ionic charge $q_i^{eff}(t)$ is the bare ionic charge reduced by the local number of quasi-free electrons per ion[28] and ii) the unshielded case, where the effective charge corresponds to the bare charge of the ion $q_i^{eff}(t) = q_i(t)$ and the surrounding electrons are completely neglected.

The X-ray yield per cluster is determined by keeping track of the K – shell vacancies produced in the cluster ions by electron-impact ionization and correcting them by a mean fluorescence yield $<\omega_k>$. For a comparison with the experimental yields, the Gaussian spatial intensity profile of the laser beam is taken into account by averaging over several laser intensities for each data point. The fraction of the gas flow condensing to clusters, in the following referred to as clustering fraction $\eta$, is not well-known experimentally but enters the theoretical determination of absolute X-ray yields. Estimates for the clustering fraction will be given below.

## III. DEPENDENCE OF THE X-RAY YIELD ON THE LASER INTENSITY

### A) Experimental results

The experimental absolute X-ray yield dependence on laser peak intensity for a typical short pulse duration of $\tau = 60$ fs (FWHM) and large argon clusters with $<N> = 5.1 \times 10^5$ atoms per cluster (fig. 1) resembles the behavior previously observed for different cluster sizes[4,17,18] as well as for xenon clusters[19,34]: a rapid onset of the X-ray yield above a well defined threshold $I_{peak} = I_{th}$ followed by a slower increase accurately described by a $I_{peak}^{3/2}$ law. Above the threshold, the dependence of the X-ray yield can be accurately fitted, over 4 orders of magnitude, by the time-average of the effective focal volume $V_{eff\,foc}$. $V_{eff\,foc}$ corresponds to the volume in which, for a given intensity I in the laser focus, the local laser intensity exceeds the threshold intensity $I_{th}$ [35]:

$$V_{eff.foc.} = \frac{\left(\pi w_0^2\right)^2}{\lambda} \left\{ \frac{4}{3} \cdot \left(\frac{I}{I_{th}} - 1\right)^{1/2} + \frac{2}{9} \cdot \left(\frac{I}{I_{th}} - 1\right)^{3/2} - \frac{4}{3} \cdot arctg\left(\frac{I}{I_{th}} - 1\right)^{1/2} \right\} \quad (2)$$

where $w_0$ stands for the radius at $1/e^2$ of the Gaussian intensity profile of the laser beam. The laser intensity threshold deduced from fitting eq. (2) to the data of figure 1 is found to be $I_{th} \approx 1.4 \times 10^{15}$ W/cm². In complete agreement with our previous results[17,19] the maximum free electron oscillation energy in the laser field $2U_P = F_L^2/(2\omega^2) \approx 170$ eV at the threshold intensity $I_{th}$ is more than one order of magnitude below the K-shell binding energy of argon ($E_K = 3.2$ keV for neutral argon and 4.1 keV for helium like argon ion). Clearly, the laser electric field alone cannot be responsible for heating the electrons to energies large enough to create inner-shell vacancies and the resulting K X-ray emission.

This close link between the effective focal volume and the total yield is well known in laser-atom interactions for the Optical Field Ionization process[36], and has recently been taken into account to explain ionic emission from silver clusters embedded in helium nanodroplets[37]. The effective focal volume is proportional to the number of clusters experiencing a laser field with $I > I_{th}$ and thus participating in the X-ray emission. The good match between the yield and $V_{eff\,foc}$ suggests that the X-ray emission probability is only weakly[21] dependent on the laser intensity once the threshold value $I_{th}$ has been reached. The mean X-ray emission probability per cluster atom $<P_X>$ (averaged over all positions in the cluster and in the laser volume) can be estimated from the total X-ray yield $N_x$ and the number of cluster atoms in $V_{eff\,foc}$,

$$<P_X> = \frac{N_x}{n_{mean} V_{eff\,foc} \eta} \quad (3)$$

with $n_{mean}$, the mean atomic density of the gaseous jet and $\eta$, the clustering fraction. $\eta$ depends strongly on the thermodynamical and geometrical characteristics of the cluster jet and, as mentioned before, is not well known. Nevertheless, the properties of our cluster jet can be estimated from experimental studies[25] with similar nozzles supported by simulations of gas flow[24]. For the present experimental conditions a fraction of beam particles condensing in clusters of $\eta \approx 0.33$ was found. Eq. (3) gives a mean X-ray emission probability $<P>$ in the range of $1 - 2 \times 10^{-5}$ per cluster atom.

Using a mean fluorescence yield $<\omega_k>$ of 0.25 corresponding to the experimentally observed mean charge states (see below) and the clustering fraction $\eta \approx 0.33$, we can evaluate the total absolute X-ray yield from the simulation without any freely adjustable parameter (squares in fig. 1). Beside the simplifications underlying the theoretical approach, the quantitative agreement between the ab – initio simulations and the experimental results over 4 orders of magnitude in the absolute X-ray yield is remarkable. In particular, the behavior of the absolute X–ray yield close to the intensity threshold is well reproduced.

The weakly varying X-ray emission probability with the laser intensity hints at an important avenue for optimizing the X-ray yield for potential applications. Once the threshold intensity is exceeded, the signal grows with the number of clusters contained in the effective focal volume. Consequently, the X-ray yield can become larger by increasing the laser beam waist at fixed laser energy. The effective focal volume as a function of the beam waist (fig. 2) displays a well pronounced maximum indicating the optimum setting for maximizing the X-ray yield. For a fixed energy per pulse of 50 mJ and a pulse duration of 60 fs, an increase of the half-width at $1/e^2$ in intensity ($w_0$) from 15 µm to 80 µm results in a gain of more than a factor of 7 in the absolute X-ray yields. Therefore, defocusing of the laser beam may allow a significant increase of the number of emitted X-rays, provided the overlap of the laser spot and the cluster jet is well controlled and the reduced intensity is still above the threshold value. This trend could also explain the recent experimental result obtained by Chen et al. [38], apart from propagation effects since a high density medium (i.e.>$10^{18}$ at./cm$^3$) is used. They observed a displacement of the jet-nozzle relative to the laser focus by around 2 mm to be optimal for maximizing the Kα photon flux when imaging biological specimens with a femtosecond-laser-driven X-ray source using argon clusters.

The variation of the charge state distribution with the laser intensity has also been investigated in detail by high resolution spectroscopy for the same experimental conditions as figure 1. Two typical spectra recorded by the crystal spectrometer for different laser intensities are displayed in figures 3b and 3c. Characteristic argon Kα lines (i.e. transitions : $1s2pn\ell \rightarrow 1s^2n\ell$) are resolved, allowing the identification of the different charge states between 12+ and 16+ with the same Bragg angle position of the crystal. It is worth noting that integration over the whole charge state distribution leads to total K X-ray yields that can be compared to the data obtained from the semiconductor detectors (shown in figure 1). The relative intensity of the different Kα lines gives information on the L- (and M-) shell filling of the emitting argon ions and allows precise evaluation of a mean X-ray energy which is related to a mean ion charge state (fig. 3a). A saturation of the charge state distributions towards the spectrum 3c has been found for laser intensities that exceed 3-4 times the threshold intensity, corresponding to a constant mean X – ray energy of 3096.6±1.6 eV. In contrast, close to the intensity threshold value at $2.1 \times 10^{15}$ W/cm$^2$, even if the mean energy of the high resolution spectrum is only by 16 eV lower (3083±2 eV), the charge state distribution is strongly modified (see spectrum 3b) as a consequence of a progressive extinction of the highest charge states.

To compare with theory, the experimental charge state distribution of ions with a K-shell vacancy can be extracted from the high resolution spectrum of figure 3c, by taking into account, for each charge state, the corresponding mean fluorescence rate[39] $\omega_K^{q+}$ weighted over all electronic configurations up to the 2p state (*i.e.* all $1s2s^n2p^m$ configurations). The result is compared with the theoretical charge-state distribution averaged over the spatial intensity distribution of the laser beam (fig. 4). The high ionic charge state and the distribution width are rather well reproduced for the high intensity regime (I > $10^{16}$ W/cm$^2$). The experimental and theoretical results differ only by one charge state. For lower laser intensities near the threshold, our simulation systematically underestimates the mean experimental charge state by more than 3 units, i.e. it fails to efficiently deplete the L-shell. We note that a different simulation method[40] gives charge state distributions similar to ours. The origin of this discrepancy to the experimental data is not yet well understood.

### B) Single cluster dynamics

Our theoretical model reproduces the unexpectedly low intensity threshold for X-ray production. For a better understanding of the heating mechanism underlying the efficient X-ray production, we describe the dynamics of a single cluster at a peak laser intensity of $I = 7.2 \times 10^{15}$ W/cm$^2$ in more detail.

As soon as the laser pulse reaches the threshold intensity for over-barrier ionization (OBI) of neutral argon atoms ($I_{OBI} = 2.2 \times 10^{14}$ W/cm$^2$), a plasma with electron density $\rho^{(e)}(t_{OBI}) = 2.7 \times 10^{22}$ cm$^{-3}$ is formed. The number of electrons rapidly increases due to efficient further ionization of the cluster ions. The dominant process is electron-impact ionization, with field-ionization being only efficient in the beginning of the pulse at the cluster surface.

At low intensities close to threshold and short pulse durations, large angle collisions in the presence of the laser field has been found to significantly contribute to effective heating[18]. At higher laser intensities, two effects mainly determine the electron dynamics. Firstly, the electron cloud is collectively driven by the laser field with respect to the ionic background and the cluster behaves similarly to a polarizable sphere[26]. Due to the high electronic density, the associated plasma frequency $\omega_P^2 = 4\pi\rho^{(e)}(t)/3$ exceeds the driving frequency of the laser and the polarization of the cluster results in a screening of the laser field inside the cluster. On the cluster poles, however, the displacement of the electron cloud results in unbalanced charges that can enhance the electric field. The second effect is the charging of the cluster (outer ionization): a fraction of the quasi-free electrons leaves the cluster during the laser pulse, resulting in the build-up of an overall positive charge on the cluster surface. The combination of these two effects, charging of the cluster and polarization of the cluster, leads to a strong asymmetry of the electric field $F_z(z)$ along the z-axis (fig. 5a). At the left cluster pole, the positive background charge is almost completely compensated by the displaced electron cloud (fig. 5b), thus suppressing the electric field, while on the right pole, a sheath region with a large number of unbalanced ionic charges and a strongly enhanced electric field arises. Half a laser cycle later, this charged sheath region has disappeared at the right pole while a similar sheath has build up at the left side of the cluster, thus interchanging the roles of the two cluster poles. It is worth noting, that $F_z(z)$ shows no sign of plasma resonances where the local electronic plasma frequency matches the laser frequency. Similar charge-density and field distributions have also been observed with PIC simulations at very high intensities above $10^{17}$ W/cm$^2$ [41].

The influence of this periodic build-up and collapse of sheath regions on the dynamics of the electrons can best be understood when examining the electronic phase space. The phase space projection $(z,v_z)$ for a subset of electrons within $R < 2.5$ nm of the z-axis (fig. 6) features an elongated high-density strip (gray in fig. 6) along the horizontal axis that represents the vast majority of the cluster electrons. Due to the small effective electric field inside the cluster and the ongoing production of slow electrons by sequential ionization, the electrons inside the cluster have, on average, kinetic energies well below the ponderomotive energy $U_P = 430$ eV of the free laser field. Almost the entire energy distribution can be well described by Maxwell-Boltzmann functions with time-dependent temperatures that do not exceed 170 eV during the entire duration $2\tau = 120$ fs of the laser pulse. However, the fraction of fast electrons reaching energies high enough to produce K-shell vacancies in the cluster ions, are grossly underestimated by the Maxwell-

Boltzmann distributions[26]. The fast electrons (blue in fig. 6) are heated in the sheath regions at the cluster poles (red arrows in fig. 6): when the sheath collapses at the pole a small fraction of the slow electron population can leak out of the cluster (fig. 6a). As soon as the sheath builds up again, these electrons are strongly accelerated back into the cluster (fig. 6b) and travel through the cluster as a front of fast electrons (fig. 6c). Again, the situation at the poles is inversed after half a laser period (compare green and red arrows in 6a and 6c). Similar heating mechanisms have also been discussed for laser pulses impinging on solid surfaces[42] and for capacitively coupled radio-frequency discharges[43]. In a 2D PIC code where the cluster is represented by an infinitely long cylinder[15,44], the authors suggest a resonant heating mechanism for the production of fast electrons which, after transiting the cluster during half a laser cycle, reemerge at the other pole in phase with the laser. For long pulse durations, several bunches of fast electrons transiting through the cluster were observed and interpreted as higher order resonances. However, for all cluster sizes and laser parameters we have examined, we could not observe any clear signature of such resonance effects for our spherical clusters. Indeed the complex spatial and temporal evolution of the electric field and the broad velocity distribution of the fast electrons (fig. 6) make the formation of a sharp resonance rather unlikely. In fact a large fraction of the fast electrons crosses the cluster several times, without however increasing on average their energy significantly from one passage to the next. The analysis of the phase space projection also explains the perturbation of the charge neutrality observed inside the cluster (fig. 5): the transiting front of fast electrons (blue arrow in fig. 6a) disturbs the colder background plasma in the cluster and a wake field oscillating at the local plasma frequency is formed behind the electron front (fig. 5).

The dynamics of the fast electrons is also reflected in the spatial distribution of the ions with K-shell vacancies, which are preferentially situated in proximity of the laser polarization axis. We note that even though the fast electrons[45] as well as the explosion dynamics of the ions[46] carry the signature of the laser polarization, the experimentally observed X-rays are emitted isotropically[3,21]. This is to be expected, since the X-ray emission occurs from many atomic configurations which are strongly mixed by the perturbing nanoplasma environment.

Additional insights into the heating mechanism and its link to the production of X-rays can be gained when examining the maximum kinetic energy $E_{max}$ of electrons inside the cluster (fig. 7). The time dependence of $E_{max}$ is governed by the evolution of the electrostatic potential $\Phi(\mathbf{r}=0,t)$ at the center of the cluster, which gives the maximum kinetic energy gained from the monopole field produced by the charging of the cluster. This estimate can be improved by additionally taking into account the maximum energy $F_{osc}^2/(2\omega^2)$ electrons can get from the acceleration in the oscillating field $F_{osc}$ created at the cluster poles by the laser and the cluster polarization. The production of X-rays sets in as soon as the fastest electrons in the cluster reach the threshold energy necessary for producing a vacancy in the K-shell of a moderately charged argon ion (q<8) by electron impact. For laser intensities close to the observed threshold $I=1.4\times10^{15}$ W/cm$^2$, the maximum energy only barely exceeds $E_K$. For intensities below the threshold ($I=4\times10^{14}$ W/cm$^2$), the charging of the cluster is not sufficient for the electrons to reach $E_K$. We can conclude that $\tilde{E}_{max}(t) = \Phi(\mathbf{r}=0,t) + F_{osc}^2/(2\omega^2) = E_K$ predicts reliably the threshold for the production of K-shell vacancies and, in turn, K X-ray emission.

## IV. INFLUENCE OF THE LASER PULSE DURATION

### A) Evolution of the X-ray production threshold

Since K-shell vacancy production sets in as soon as the cluster is sufficiently charged to allow electrons to be heated beyond $E_K = 3.2$ keV, longer pulse durations at a given peak intensity should enhance X-ray production. One can therefore expect a lowering of the intensity threshold for the production of X-rays with increasing laser pulse duration. Indeed, the experimentally observed intensity threshold for K – shell vacancy production decreases when the pulse duration increases (fig. 8) from $I_{th} = 2.9 \times 10^{15}$ W/cm$^2$ for a pulse duration of $\tau = 55$ fs to $I_{th} = 2.8 \times 10^{14}$ W/cm$^2$ for $\tau = 570$ fs. The latter corresponds to a threshold reduction by a factor 10 and is close to the intensity $I_{OBI}$ [Ar] $= 2.2 \times 10^{14}$ W/cm$^2$ necessary to produce the first quasi-free electrons by over-barrier ionization of the argon atom[35]. It is worth mentioning that for all pulse lengths the behavior of the x – ray yield is again well fitted over 5 orders of magnitude by the raise of the effective focal volume (eq. 2).

The experimental data (fig. 8) were obtained with settings of the cluster source different from those presented in figures 1 and 3: a wider cluster beam (8 mm instead of 4 mm FWHM), a mean cluster size of approximately $<N> = 3.7 \times 10^4$ atoms per cluster and a mean atomic density $n_{mean}$ more than a factor 10 smaller. The simulation results in fig. 8 assume now a reduced clustering fraction of $\eta = 0.06$ and a mean fluorescence yield $<\omega_k>$ of 0.12. No independent experimental data on $\eta$ are presently available. Note, however, that the assumed clustering fraction provides only one overall renormalization factor for all three curves, but does distort neither the comparison for the relative yield change as a function of I or $\tau$, nor the position of the thresholds.

The simulation results reveal that the heating in the sheath region is efficient enough to account for X-ray production even at laser intensities where the ponderomotive energy of free electrons driven by the laser field is as small as 10 eV. The simulation agrees with the observed absolute X-ray yields for $\tau = 55$ fs and $\tau = 140$ fs quite well. For the longest pulses with $\tau = 570$ fs, the simulation overestimates the absolute experimental X-ray yields by approximately a factor 3 but, nevertheless, reproduces the intensity dependence.

The strong dependence of the X-ray yield on the pulse duration points to a strong influence of the ion dynamics on the electron dynamics. As the cluster expands, a density gradient of the positive background charge is formed enlarging the sheath region where efficient electron acceleration takes place, and thus boosting the X-ray production. Errors in the speed of the ion expansion velocities may cause increasingly inaccurate estimates for the electron acceleration and K-shell vacancy production for long pulse durations. In the present simulation, the local electronic charge distribution is not resolved within the discretized cells and thus contributes to shielding of ions only via the mean field. Consequently, the effective shielding of the repulsive ion-ion interaction driving the cluster expansion is not accurately represented. In order to estimate the size of possible errors and its effect on K-vacancy production we consider in the following two limiting cases: i) complete shielding of the ion by reducing its instantaneous charge $q_i(t)$ to an effective charge $q_i^{eff}(t)$ determined by the local electron density in the cell (see section 2B), ii) the complementary case which uses the bare ionic charge $q_i(t)$ for the ion dynamics, i.e. neglects all shielding effects caused by the local electron density. In the first case, the ion-ion repulsion is underestimated and the Coulomb explosion is delayed. In the second case, the cluster expansion is likely to be overestimated.

The number of K-shell vacancies obtained for a single cluster at different laser parameters in these two limiting cases is compared in figure 9. For relatively short pulses ($\tau = 55$ fs) during which only the onset of expansion occurs, the evolution of the K-shell ionization probability per atom is not very sensitive to the assumptions about the effective shielding. For longer pulses, shielding corrections become important. For weak laser intensities, neglecting shielding leads to lower threshold intensities $I_{th}$ than with shielding included and than observed in the experiment. Conversely, for high laser intensities, the observed X-ray yield is overestimated by the simulation with shielding and better agreement is achieved with the simulation neglecting shielding. The origin of the enhanced production of fast electrons lies in the extended sheath region in the early stages of the expanding cluster. The heating is terminated when the disintegration is well advanced leading to a saturation of the K-shell production. This situation suggests an intensity dependent screening length $l_{sc}(I)$ which, in turn, can be interpreted as a temperature dependent screening length $l_{sc}(T_e)$. This conforms with the intuitive picture, that in strong laser fields the more energetic ("hotter") electrons are less influenced by the ionic potentials and shield the ions less efficiently. For weaker laser intensities, i.e. close to the threshold, stronger electron-ion correlation persists. An improvement to the present calculation would be a Debye-like shielding description taking the kinetic energy of the heated electrons in the cell into account.

It should be emphasized that the dependence of hot electron production on pulse duration and the cluster expansion is unrelated to nanoplasma resonances. The critical electron density for a nanoplasma resonance is only reached towards the very end of the cluster explosion when the ionic density is by far too low for efficient K-shell vacancy production by electron-ion collisions.

### B) Evolution of the charge state distribution with the pulse duration

While the X-ray production is controlled by the fraction of fast electrons, the ionic charge-state distributions are governed by the global electron energy distribution that contributes to the ionization of M- and L-shells. The experimental charge-state distributions (fig. 10) have been measured in the saturation region of high laser intensities for three different laser pulse durations. For these considerably smaller clusters (factor 10 fewer atoms than in those in fig. 4) the charge state distribution is centered at lower charges. The theoretical charge-state distributions are obtained by averaging the results from single clusters over the effective focal volume (i.e. from $I_{th}$ to $I_{peak}$).

The experimental distributions are best reproduced by the simulation neglecting the ionic shielding, which reproduces well the slight shift towards higher charge states as the pulses grow longer. With shielding included, however, the resulting charge states for the longest pulses are far too high.

### C) Evolution of the X-ray yield with the pulse duration at constant laser energy

Since the K X-ray emission is influenced by both the laser peak intensity $I_{peak}$ and the pulse duration $\tau$, systematic investigations of the dependence on $\tau$ require the simultaneous control of the intensity. One possibility is to keep the laser energy, $E \propto$

($I_{peak} \times \tau$), constant at values such that the intensity threshold $I_{th}(\tau)$ is exceeded for all studied pulse durations.

Due to the spatial laser intensity profile, the X-rays are produced by an ensemble of clusters subjected to laser intensities from $I_{th}(\tau)$ to $I_{peak}(\tau)$. Because of the uncertainty in the description of the cluster expansion, we restrict ourselves to qualitative predictions for the dependence of the total X-ray yield $N_X(\tau)$ on the laser pulse duration. We expect our model without ion shielding to give better results at high laser intensities and therefore employ it to predict the probability $P_K$ to produce K-shell vacancies. $P_K$ at high laser intensities is found to grow with increasing pulse lengths before saturating due to cluster disintegration (fig. 11a). The other ingredient for the calculated X-ray yield is the effective focal volume $V_{eff\,foc}$ which is proportional to the number of clusters participating in the X-ray production, and is governed by the threshold intensity for X-ray production $I_{th}(\tau)$ and the peak intensity $I_{peak}(\tau)$ (eq. (2)). To predict the dependence $I_{th}(\tau)$, ion shielding is included since this model is expected to be more accurate at low laser intensities. The resulting $V_{eff\,foc}$ shrinks with increasing pulse lengths (fig. 11b). As $N_X(\tau) \propto P_K(\tau) \times V_{eff\,foc}(\tau)$, the increase of the vacancy production per cluster counteracts the decrease of the number of participating clusters. As a result of this competition, the total X-ray yield goes through a maximum at the optimal laser pulse duration $\tau_{max} \sim 140$ fs (fig 11c.).

Such an optimal laser pulse duration for the X-ray emission has indeed been observed experimentally for large argon clusters with <N> between $5.1 \times 10^5$ and $1.8 \times 10^6$ atoms irradiated by laser pulses of constant energy E = 20 mJ (fig. 12). The qualitative behavior predicted theoretically for smaller clusters (fig. 11) is well reproduced: after a clear increase by more than a factor 2 between 50 and 140 fs, the X-ray yield decreases from 140 to 800 fs. The optimal pulse duration (140 fs) is close to the theoretically predicted $\tau_{max}$. The presence of an optimal pulse duration and its dependence on the cluster size has often been used as an indication for the appearance of a nanoplasma resonance[13,14]. However, in the size range studied here, we do not observe any significant shift of this optimum when varying the cluster size. Therefore, under the present conditions, the competition between the number of emitting clusters and the number of emitted X-rays per cluster seems to be the appropriate explanation for the observed pulse length dependence.

### D) Dependence of the X-ray yield on the pulse duration at constant laser intensity

An alternative approach to quantifying the X-ray yield as a function of the pulse duration is keeping the laser peak intensity constant, which implies increasing pulse energies for longer pulses. After an initial strong increase by more than one order of magnitude in the X–ray yield between 55 fs and 140 fs (fig. 13), we observe a slow saturation for longer pulse durations. These results can again be understood when considering that the X-ray yield $N_X \propto V_{eff\,foc} \times P_K$ is governed by both the effective focal volume and the probability for K-shell vacancy production in a single cluster. $P_K$ at constant laser intensity increases with increasing pulse durations, saturating for long pulses due to cluster disintegration (see fig. 9). As $I_{peak}$ is fixed, $V_{eff\,foc}(\tau)$ is governed by $I_{th}(\tau)$ alone. The drop in the threshold intensity for longer pulses (see fig. 8) results in an increasing $V_{eff\,foc}(\tau)$. For very long pulses the intensity threshold reaches its minimum value close to $10^{14}$ W/cm$^2$ and $V_{eff\,foc}(\tau)$ no longer changes. At constant laser peak intensity both $V_{\text{eff foc}}(\tau)$ and $P_K(\tau)$ initially increase and then saturate when going to longer pulse durations. Therefore, the absolute X-ray yield, given by the product $V_{\text{eff foc}} \times P_K$, displays the same trend (see fig. 13).

## V. CONCLUSION

We have studied, experimentally as well as theoretically, the influence of laser intensity and pulse duration on the X-ray emission from argon clusters exposed to intense and short laser pulses. The absolute X –ray yield is found to be governed not only by the collision dynamics in a single cluster but also by the variation of the effective focal volume with the laser intensity. The effective focal volume is proportional to the number of clusters experiencing laser intensities that exceed the intensity threshold for X – ray production. Using a mean field approach which includes many-particle effects via Monte Carlo events, efficient heating of a subensemble of electrons up to energies in the keV range is clearly identified. Strongly enhanced electric fields in sheath regions formed at the cluster poles by the combined action of cluster charging and polarization are found to cause the production of these fast electrons, even at low laser intensities $I < 10^{15}$ W/cm$^2$. The absolute X-ray yields as a function of laser intensity as well as the dependence of the intensity thresholds $I_{th}$ on the pulse duration are quantitatively reproduced by our simulation. Furthermore, at high laser intensities, a good agreement is found between the experimental and theoretical charge state distributions of ions with a K – shell vacancy. Their high charge states up to $Ar^{16+}$ reflect a very efficient stripping of the L-shell.

The competition between electron heating mechanisms and ionic motion (i.e. the cluster expansion) is investigated by varying the laser pulse duration. The intensity threshold for X-ray production drops for long pulses down to the threshold intensity $I_{OBI}$ for over-barrier ionization of neutral argon. For relatively short pulse lengths ($\tau$ up to ~100 fs), the K-shell ionization probability is rather insensitive to the dynamics of the ions, while for longer pulses, the expansion of the cluster plays a significant role. For an accurate theoretical description of the ionic motion, the shielding of the ions by slow quasi-free electrons has to be taken into account. While a full molecular dynamics approach that includes all electron-ion correlations seems not feasible for the present cluster sizes, an improved Debye-like shielding description appears to be in reach.

Finally, we observe an optimum pulse duration at constant laser energy for which the total X-ray yield displays a maximum. As this optimum is not influenced by the cluster size, we rule out a nanoplasma resonance effect. Instead we explain the existence of such an optimum pulse duration by the competition between the effective focal volume (i.e. number of emitting clusters) and the K-shell ionization probability per cluster. In view of potential applications, the present results demonstrate that the absolute X-ray yield may be significantly increased by defocusing the laser beam and by tuning the pulse duration to the optimum duration around a few hundred femtoseconds.

## ACKNOWLEDGEMENTS
We would like to thank O. Gobert, D. Normand, M. Perdrix and P. Meynadier from the SPAM/DRECAM in Saclay for their valuable help setting up the LUCA laser facility and their fruitful advice and discussions. This work is supported by FWF SFB-16 (Austria).

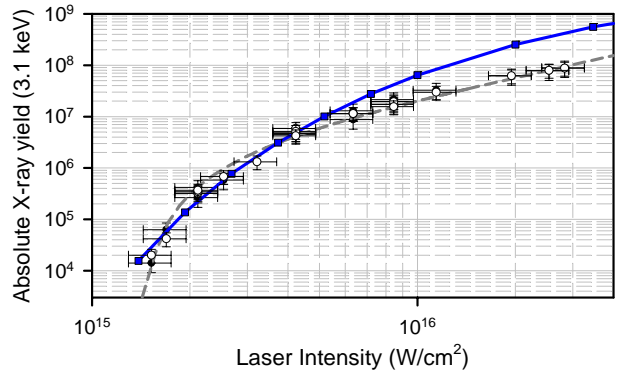

**Figure 1: Absolute 3.1 keV photon yield as a function of the laser peak intensity for $\lambda=800$ nm, $\tau = 60$ fs, and $\langle N \rangle = 5.1 \times 10^5$ argon atoms per cluster; full circle: data from the high resolution spectrometer, open circle: data from semi – conductor detector, dashed gray line: fit to the effective focal volume (eq. 2) with $I_{th} = 1.4 \times 10^{15}$ W/cm²; square: ab-initio simulation with a clustering fraction of 0.33 and a fluorescence yield of 0.25 (see text).**

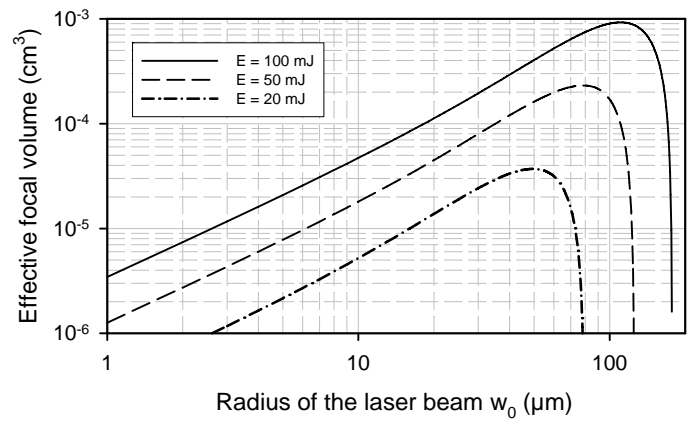

**Figure 2: Dependence of the time - averaged effective focal volume on the beam waist for different laser energies. For an intensity threshold and pulse duration of $1.4 \times 10^{15}$ W/cm² and 60 fs respectively, each $V_{eff.foc}$ displays a pronounced maximum.**

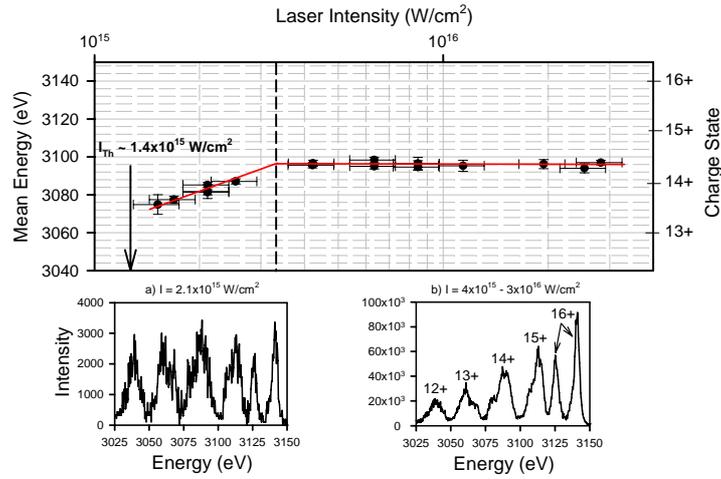

**Figure 3:** a) Laser intensity dependence of the mean energy of $1s2pn\ell \rightarrow 1s^2n\ell$ transitions from irradiation of argon clusters for the same experimental conditions as presented in figure 1. Experimental spectra obtained at b) $2.1\times10^{15}$ W/cm$^2$ and c) in the intensity range $4\times10^{15} - 3\times10^{16}$ W/cm$^2$. Data are normalized to Ar$^{14+}$.

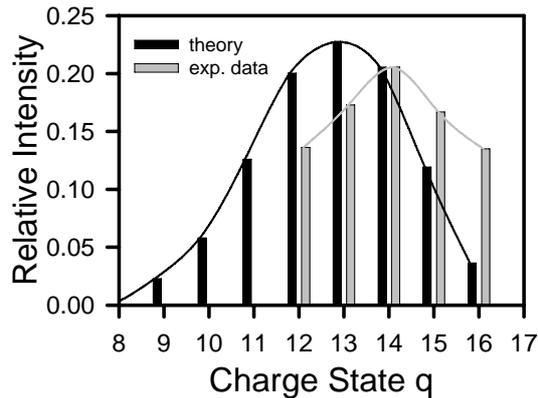

**Figure 4:** Comparison of experimental and theoretical charge state distributions obtained at $6\times10^{16}$ W/cm$^2$. The experimental distribution was obtained from spectrum 3b) assuming mean fluorescence yields $<\omega_{12+}> = 0.19$, $<\omega_{13+}> = 0.24$, $<\omega_{14+}> = 0.31$, $<\omega_{15+}> = 0.49$, $<\omega_{16+}> = 0.45$. Experimental data on charge states q < 12+ could not be extracted. Full lines to guide the eye.

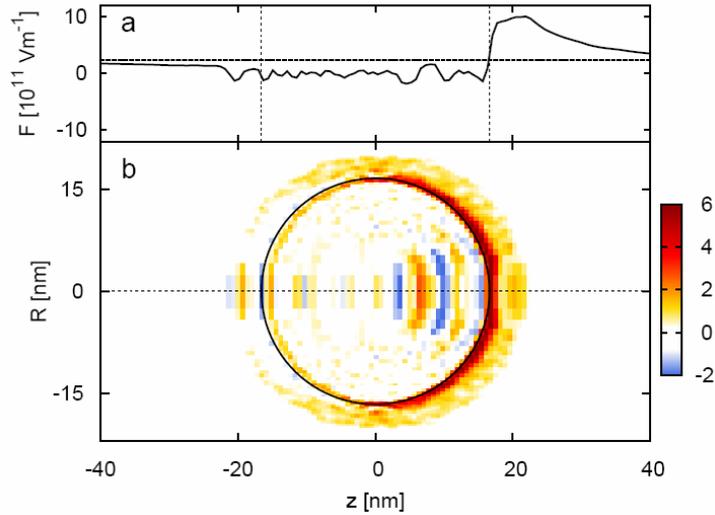

**Figure 5:** a) Snapshot of the electric field Fz(z) along the polarization axis at t=86fs where the laser field $F_L(t)=2.3\times10^{11}$ V/m (dashed line). The dotted lines mark the original cluster borders at t = 0 fs. b) Snapshot of the charge density distribution (in units of $10^9$ C/m$^3$) at the same time. The circle marks the original cluster borders at t = 0 fs. Note the sheath density near the pole on the right hand side.

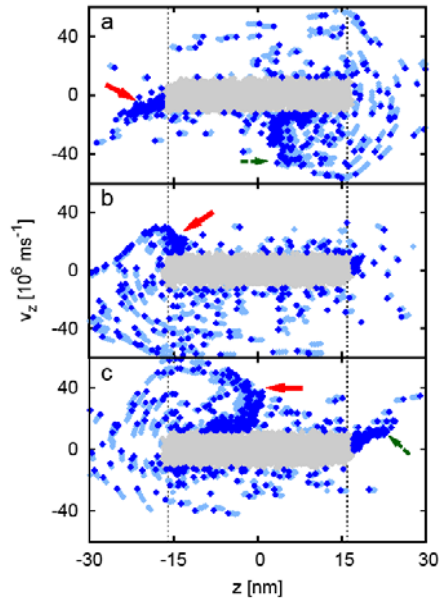

**Figure 6:** Evolution of the projected electron phase space ($z,v_z$) during half a laser cycle (a-c: t=86 fs, t=86.8 fs and t=87.5 fs). Only electrons within R < 2.5 nm of the z-axis are represented. Gray: Slow electrons inside the cluster. Blue: Electrons outside the cluster and fast electrons inside the cluster. The light-blue traces mark the path of the electrons over the previous 0.04 fs. The vertical lines show the original cluster borders. The production and evolution of fast electrons can be observed (arrows).

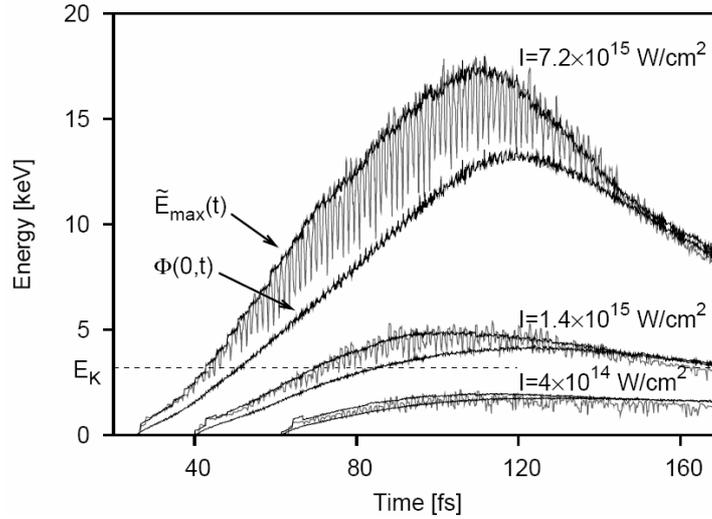

**Figure 7:** Time evolution of the maximum electron kinetic energy (gray) in the high density core of the cluster for three peak laser intensities, $I=7.2\times10^{15}$ W/cm2 (upper curve), $I=1.4\times10^{15}$ W/cm$^2$ (center curve) and $I=4\times10^{14}$ W/cm$^2$ (lower curve) and pulse duration $\tau = 60$ fs. The dashed horizontal line marks the neutral argon K-shell binding energy $E_K=3.2$ keV. Each curve is delimited below by the electrostatic potential $\Phi(r=0,t)$ at the center of the cluster and above by $\tilde{E}_{max}(t) = \Phi(r=0,t) + F_{osc}^2/(2\omega^2)$ (see text).

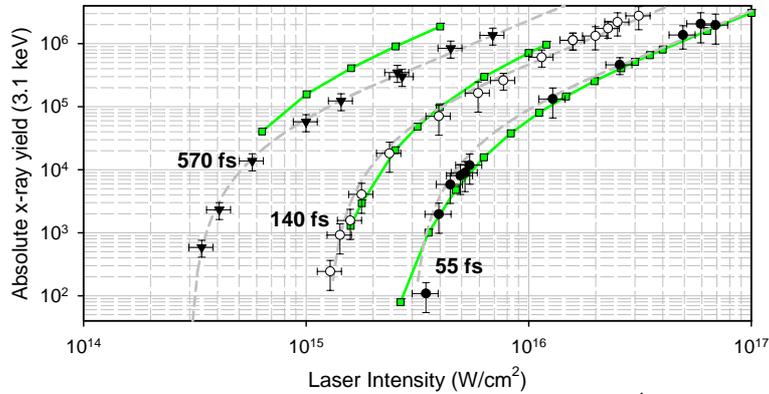

**Figure 8:** X-ray yield from argon clusters with $\langle N\rangle \sim 3.7\times10^4$ atoms as a function of laser peak intensity at 800 nm for different pulse durations of $\tau=55$, 140 and 570 fs. Gray dashed lines correspond to the fitted effective focal volume and full lines with squares to the simulation data.

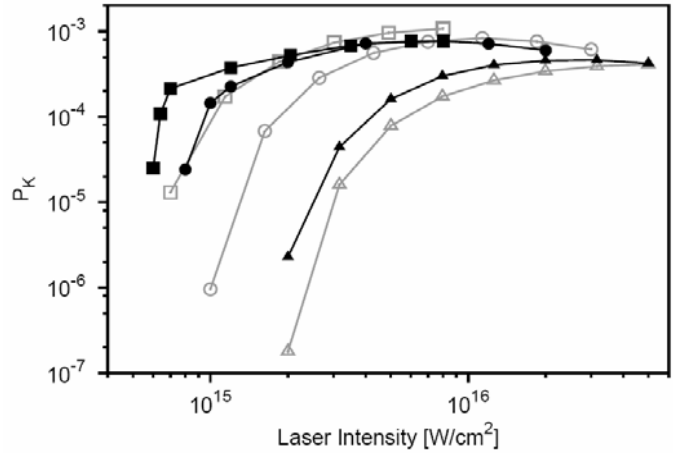

**Figure 9:** Evolution of the K-shell ionization probability per atom as a function of the laser peak intensity for different pulse durations $\tau$=55fs (triangle), 140fs (circle) and 210 fs (square) calculated with shielded ion dynamics (gray) and without ion shielding (black).

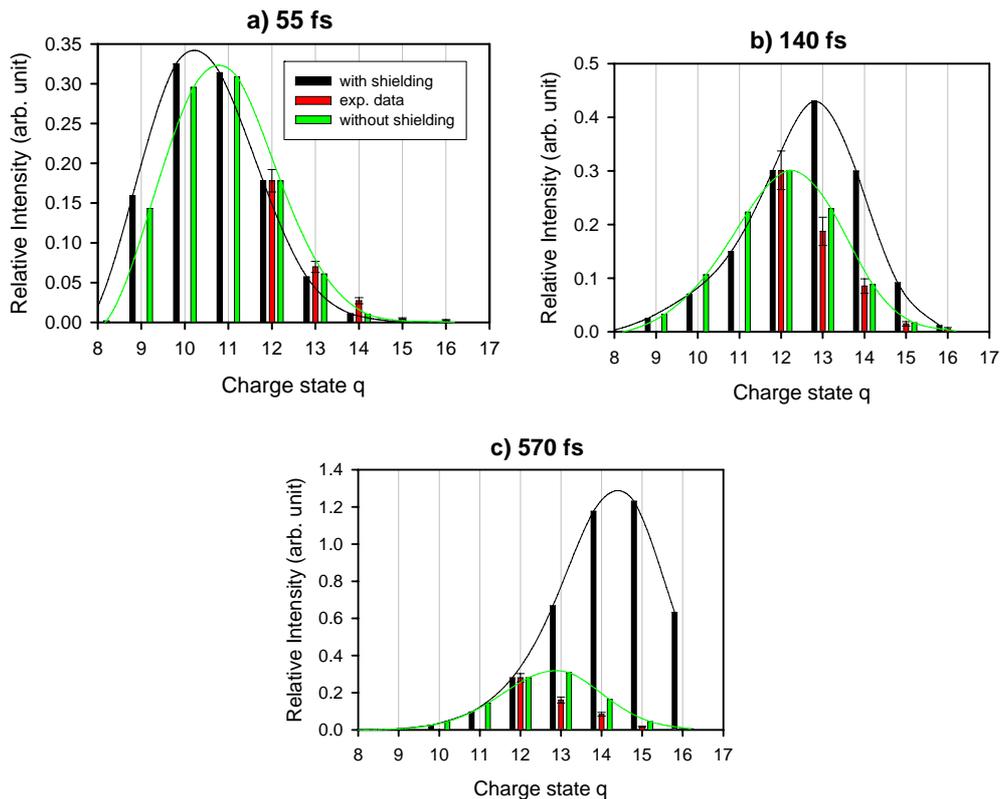

**Figure 10:** Comparison between experimental and theoretical charge state distributions for different pulse durations (same experimental conditions as figure 8, $<N> = 3.7 \times 10^4$ at./cluster), a) $\tau = 55$ fs, $I_{peak} = 4.0 \times 10^{16}$ W/cm$^2$, b) $\tau = 140$ fs – $I_{peak} = 1.6 \times 10^{16}$ W/cm$^2$ and c) $\tau = 570$ fs – $I_{peak} = 4.0 \times 10^{15}$ W/cm$^2$. Black bars: simulation with shielding, green bars: simulation without shielding (Full lines are plotted to guide eye), red bars: experimental data. All data are normalized to the lowest experimentally observed charge state (12+).

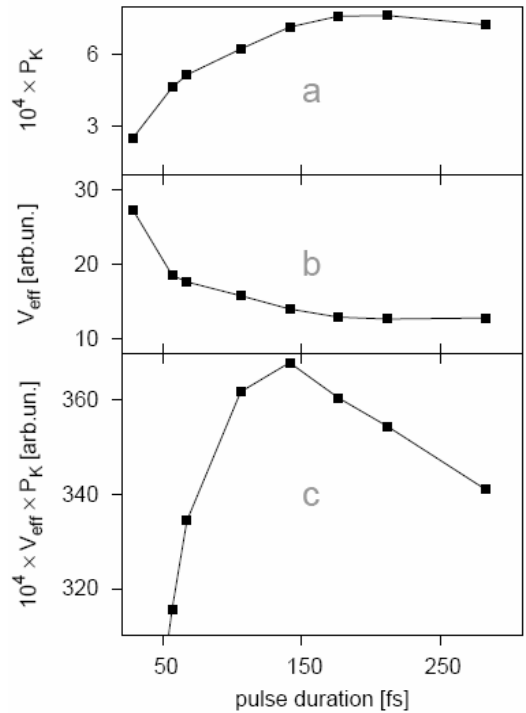

**Figure 11: Simulation of the τ dependence of X-ray yields $N_X(\tau)$ at fixed total E=20 mJ for argon clusters with $N=3.7\times10^4$ atoms: a) number of K-shell vacancies per cluster atom calculated without shielding, b) effective focal volume deduced from the threshold intensities calculated with shielding, c) qualitative estimate of the total X-ray yield $N_X$ by taking the product of a) and b).**

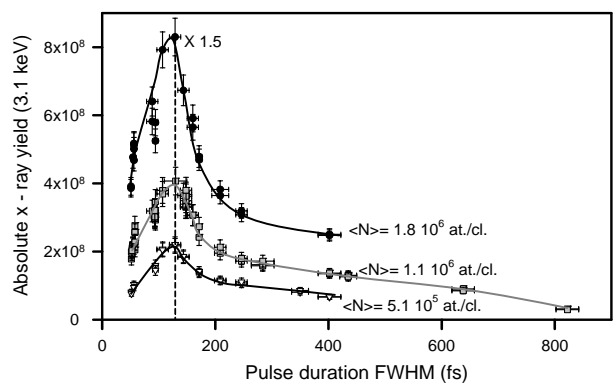

**Figure 12: Evolution of the measured absolute X-ray yield with the pulse duration at a constant energy per pulse of 20 mJ for different cluster sizes <N>. Bottom to top: $5.1\times10^5$ at./cluster, $1.1\times10^6$ at./cluster and $1.8\times10^6$ at./cluster. For sake of clarity, experimental data for $1.8\times10^6$ at./cluster are multiplied by 1.5.**

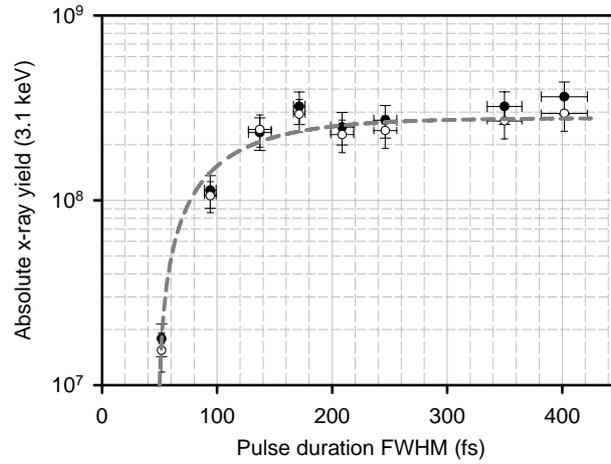

**Figure 13: Evolution of the absolute X-ray yield with the pulse duration at a constant laser intensity of $8 \times 10^{15}$ W/cm$^2$ for a cluster size $<N> = 1.1 \times 10^6$ at./cluster (full circle : data from the high resolution crystal spectrometer, open circle : data from semi – conductor detector) (lines to guide the eye).**